\pdfoutput=1
\documentclass[12pt]{article}

\usepackage{graphicx}
\usepackage[english]{babel}

\def\uhm{Department of Physics\\University of Hawaii at Manoa, Honolulu - HI (US)}
\def\Title#1{
   \begin{center}
      {\Large #1 }
   \end{center}
}
\def\Author#1{
   \begin{center}
      {\sc #1}
   \end{center}
}
\def\Address#1{
   \begin{center}
      {\it #1}
   \end{center}
}

\newenvironment{Abstract}{
   \begin{quotation}
   }
   {
   \end{quotation}
}
\newenvironment{Presented}{
   \begin{quotation}
      \begin{center}
         PRESENTED AT
      \end{center}
      \begin{center}
         \begin{large}
   }
   {
         \end{large}
      \end{center}
   \end{quotation}
}
\def\Acknowledgements{
   \begin{center}
      \begin{large}
         \bf ACKNOWLEDGEMENTS
      \end{large}
   \end{center}
}

\begin{document}

\Title{The cosmic ray electron and positron spectra measured by AMS-02}
\Author{Claudio Corti, on behalf of the AMS-02 collaboration}
\Address{\uhm}

\begin{Abstract}
The AMS-02 detector is operating on the International Space Station since May 2011. More than 30 billion events have been collected by the instrument in the first two years of data taking. A precision measurement of the positron fraction and of the positron flux in primary cosmic rays up to 350 GeV, of the electron flux up to 500 GeV and of the combined electron plus positron flux up to 700 GeV are presented. The separate and combined electron and positron fluxes are preliminary and represent work in progress. Systematic uncertainties must still be investigated further.
\end{Abstract}

\begin{Presented}
The 10th International Symposium on Cosmology and Particle Astrophysics (CosPA2013)\\
Honolulu, Hawai'i,  November 12--15, 2013
\end{Presented}

\section{The AMS-02 Detector}
The Alpha Magnetic Spectrometer, AMS-02, is a general purpose high energy particle physics detector. It was installed on the International Space Station, ISS, on 19 May 2011 to conduct a unique long duration mission ($\sim$20 years) of fundamental physics research in space.

\begin{figure}[h!]
   \centering
   \includegraphics[width=0.4\textwidth]{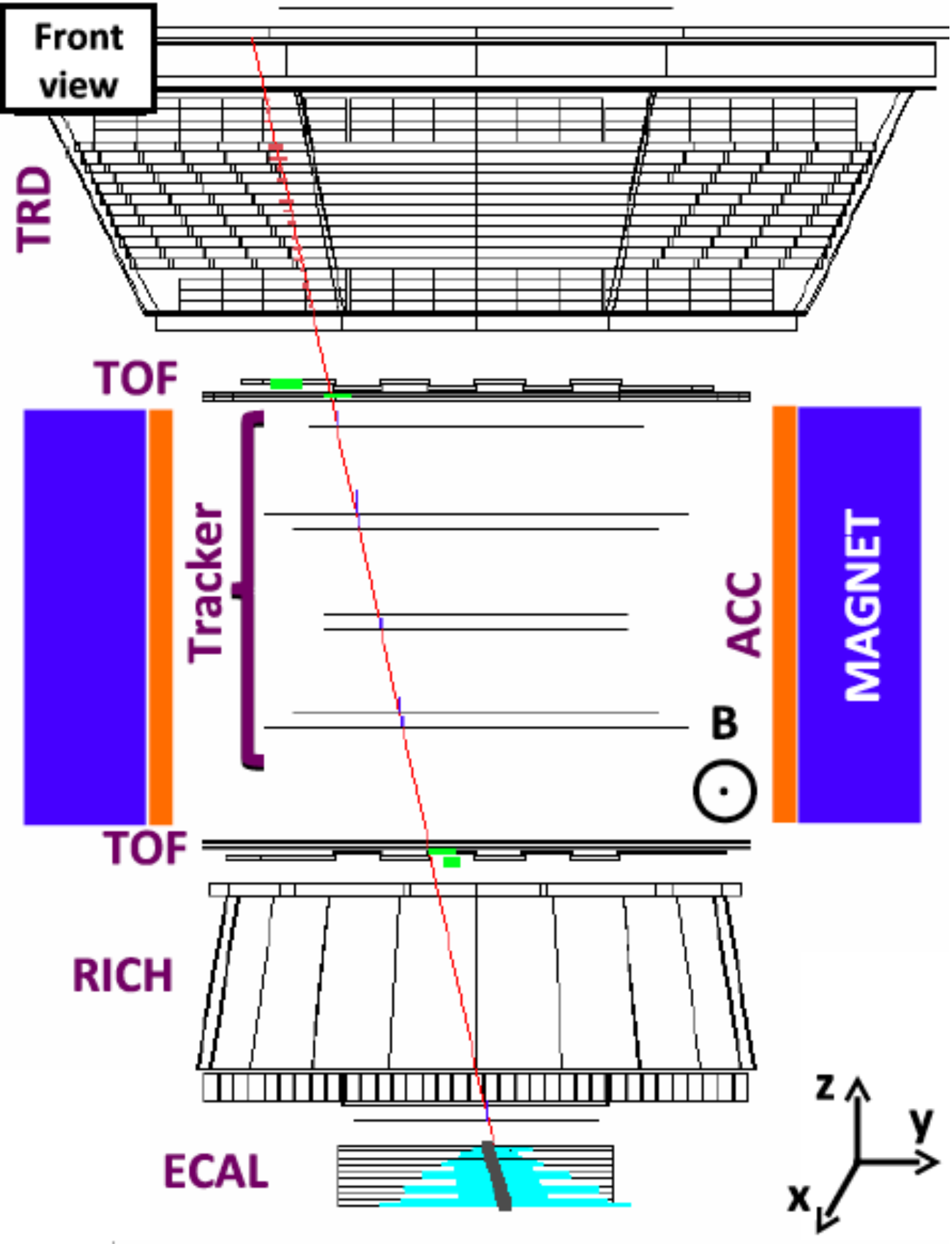}
   \caption{\emph{AMS-02 detector schematics in the event display of a 600 GeV electron.}}
   \label{fig:detector}
\end{figure}

The layout of the AMS-02 detector~\cite{bib:AMS02-PRL} is shown in Fig.~\ref{fig:detector} presenting the event display of a 600 GeV electron recorded by AMS. It consists of nine planes of precision silicon tracker, a transition radiation detector (TRD), four planes of time of flight counters (TOF), a permanent magnet, an array of anticoincidence counters (ACC) surrounding the inner tracker, a ring imaging Cherenkov detector (RICH), and an electromagnetic calorimeter (ECAL). The TRD and the ECAL are separated by the Magnet and the Tracker. This ensures that secondary particles produced in the TRD and the upper TOF planes are swept away and do not enter into the ECAL.

\begin{figure}[h!]
   \centering
   \includegraphics[width=0.9\textwidth]{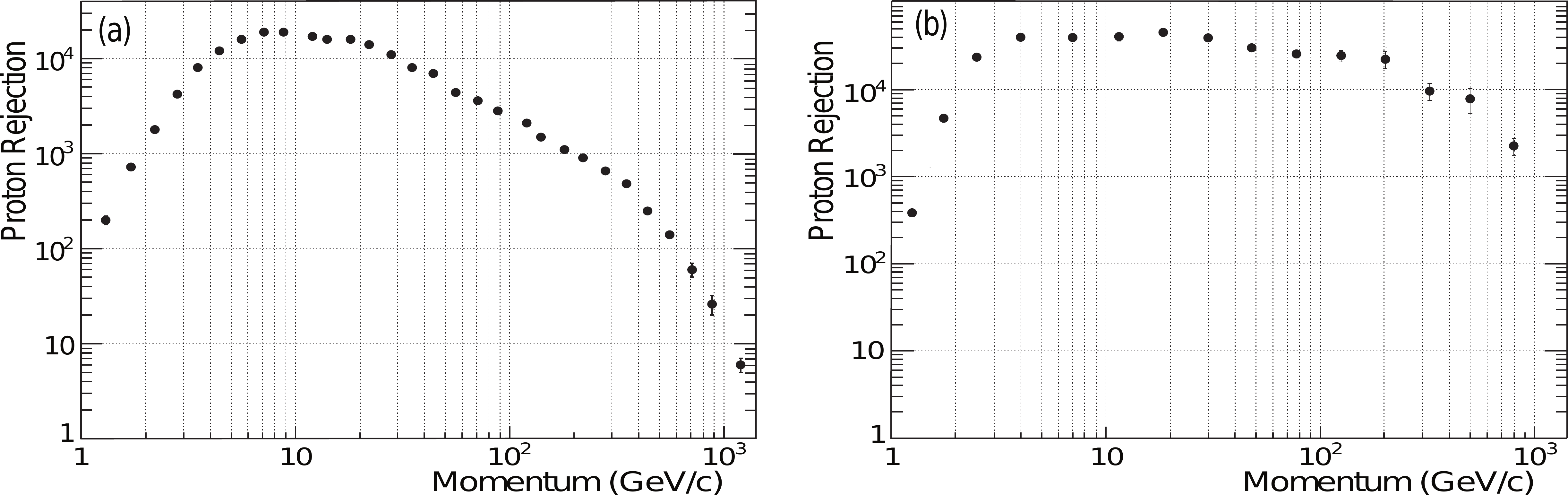}
   \caption{\emph{The measured proton rejection using the TRD} (a) \emph{and the ECAL plus the Tracker} (b) \emph{as a function of track momentum at 90\,\% selection efficiency for e$^\pm$.}}
   \label{fig:rejection}
\end{figure}

The TRD is designed to use transition radiation to distinguish between electrons and protons, and {\it dE/dx} to independently identify nuclei. It consists of 5248 proportional tubes of 6 mm diameter with a maximum length of 2 m arranged side by side in 16-tube modules. The 328 modules are mounted in 20 layers. Each layer is interleaved with a 20 mm thick fiber fleece radiator (LRP375) with a density of 0.06 $g/cm^3$. There are 12 layers of proportional tubes along the y axis located in the middle of the TRD and, along the x axis, four layers located on top and four on the bottom. The tubes are filled with a 90:10 Xe:CO$_{2}$ mixture.
In order to differentiate between electrons and protons, signals from all the TRD layers are combined in a log-likelihood probability of the electron (TRD-LLe) or proton (TRD-LLp) hypothesis. The proton rejection power of the TRD estimator at 90\,\% e$^\pm$ efficiency measured on orbit is $10^3$ to $10^4$, as shown in Fig.~\ref{fig:rejection}a.

The ECAL consists of a multilayer sandwich of lead and scintillating fibers with an active area of 648$\times$648 mm$^2$ and a thickness of 166.5 mm corresponding to 17 radiation lengths The calorimeter is composed of nine superlayers, each 18.5 mm thick. In each superlayer, the fibers run in one direction only. The 3D imaging capability of the detector is obtained by stacking alternate superlayers with fibers parallel to the x and y axes (five and four superlayers, respectively). The fibers are read out on one end by 1296 photosensors with a linearity of 1/10$^5$ per sensor. Signals from three super layers in y view (super layers 2,4,6) and in x view (super layers 1,3,5) are used in the trigger logic to select events with a shower in the calorimeter. The energy resolution of the ECAL is parametrized as a
function of energy (in GeV) $\sigma(E)/E=\sqrt{(0.104)^2/E + (0.014)^2}$. In order to cleanly identify electrons and positrons, an ECAL estimator, based on a Boosted Decision Tree, BDT, algorithm~\cite{bib06}, is constructed using the 3D shower shape in the ECAL. The proton rejection power of the ECAL estimator when combined with the energy-momentum matching requirement $E / p > 0.75$ reaches $\sim$10,000 (see Fig.~\ref{fig:rejection}b), as determined from the ISS data.
The proton rejection power can be readily improved by tightening the selection criteria with reduced e$^\pm$ efficiency.

Two planes of TOF counters are located above and two planes below the magnet. Each plane contains eight or ten scintillating paddles. Each paddle is equipped with two or three photomultiplier tubes on each end for efficient detection of traversing particles.  The average time resolution of each counter has been measured to be 160 ps, and the overall velocity ($\beta=v/c$) resolution of the system has been measured to be 4\% for $\beta\simeq 1$ and Z=1 particles which also discriminates between downward and upward-going particles. The coincidence of signals from all four TOF planes provides the charged particle trigger.

The Tracker accurately determines the trajectory and absolute charge ($Z$) of cosmic rays by multiple measurements of the coordinates and energy loss. Coordinate resolution of each plane is measured to be better than 10$\,\mu$m in the bending direction and the charge resolution is $\Delta Z \simeq 0.06$ at $Z = 1$.  Together with the Magnet, the Tracker provides a Maximum Detectable Rigidity of 2 TV on average~\cite{bib03}, over Tracker planes 1 to 9.

Specific  calibration procedures of all sub-detectors have been developed in order to guarantee the stability of the AMS-02 performances over time and no significant degradation of the apparatus has been observed during two years of operation in space~\cite{bib:icrcTracker,bib:icrcTRD,bib:icrcTOF,bib:icrcECAL}.

\section{Data sample and analysis procedure.}
Over 30 billion events have been analyzed. Optimization of all reconstruction algorithms was performed using the test beam data.

Monte Carlo simulated events are produced using a dedicated program  developed by AMS which is based on the GEANT-4.9.4 package~\cite{bib07}. This program simulates electromagnetic and hadronic interactions of particles in the materials of AMS and generates detector responses. The digitization of the signals, including those of the AMS trigger, is simulated precisely according to the measured characteristics of the electronics. The digitized signals then undergo the same reconstruction as used for the data. The Monte Carlo samples used in the present analysis have sufficient statistics sothey do not contribute to the errors.

A loose preselection is first applied to the collected events in order to keep only down going relativistic particles ($\beta > 0.8$) with  associated signals in the TRD and in the ECAL.  In order to reject particles produced by the interaction of primary cosmic rays with the atmosphere, the energy measured with the ECAL is required to exceed by a factor of 1.25 the maximal Stoermer cutoff for either a positive or a negative particle at the geomagnetic location where the particle was detected and at any angle within the AMS acceptance. Z$>$1 particles are rejected by means of the signal released in the TRD and the Tracker. This set of requirements constitutes the event pre-selection and is used in the time exposure and acceptance definitions as discussed in Sec.~\ref{sec:measurement}.

The measurement is performed in ECAL energy bins. The binning is chosen according to the energy resolution and the available statistics such that migration of the signal events to neighbouring bins has a negligible contribution to the systematic errors above 2 GeV. In each energy bin, the reference spectra of the TRD-LL estimator for electrons and protons is fitted to data varying the normalisations of the signal and the background components.

\begin{figure}[h!]
   \centering
   \includegraphics[width=0.3\textwidth]{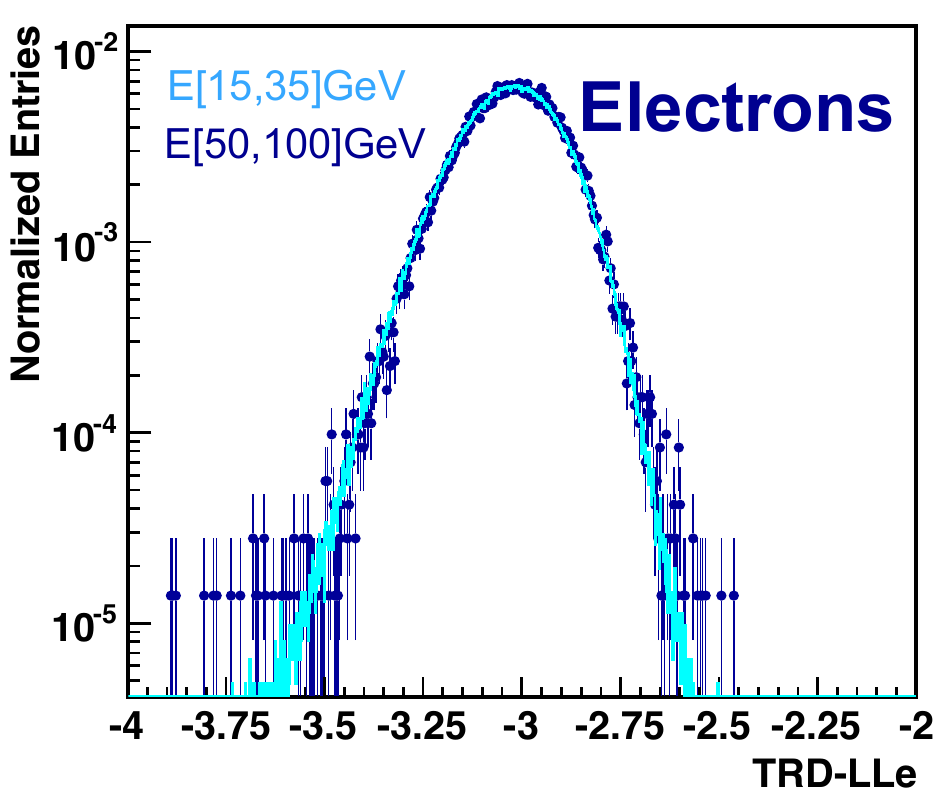}
   \ \
   \includegraphics[width=0.3\textwidth]{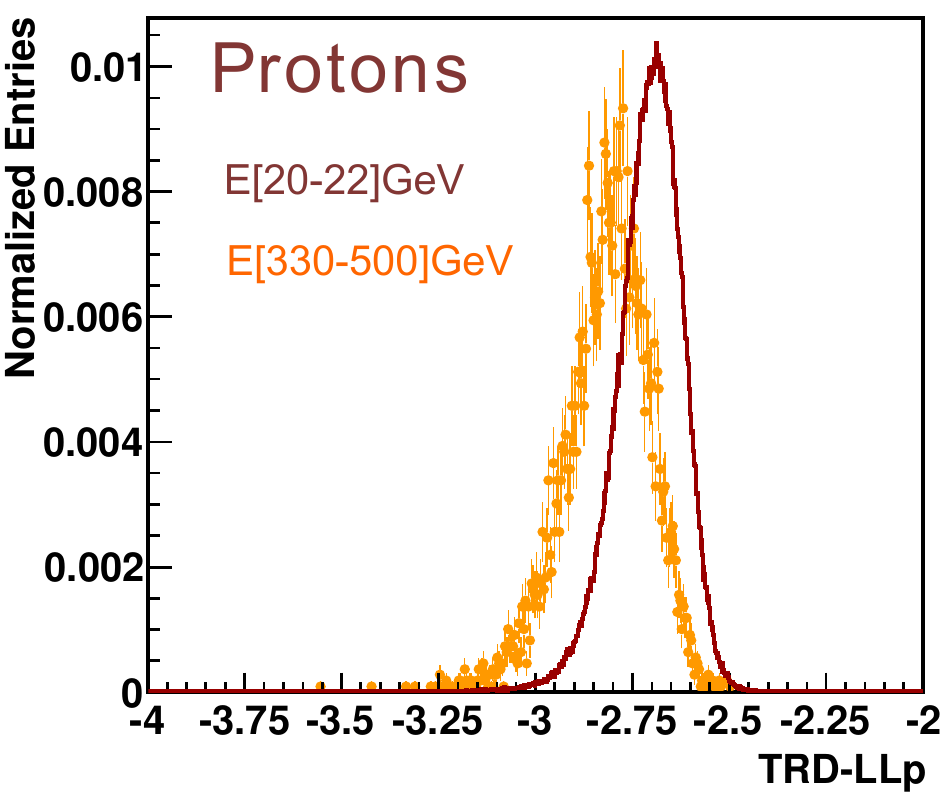}
   \ \
   \includegraphics[width=0.335\textwidth]{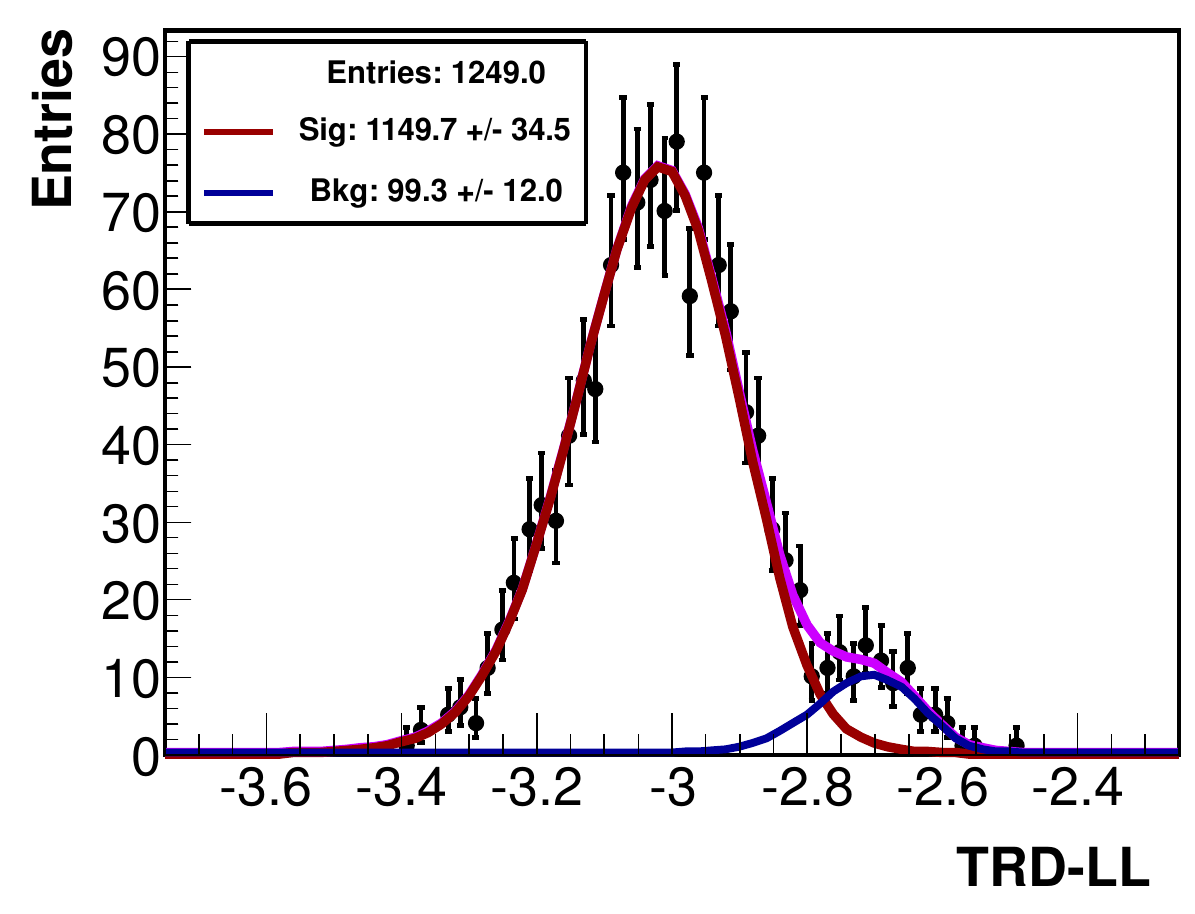}
   \caption{\emph{TRD-LL distribution for electrons} (left) \emph{and protons} (center) \emph{in different energy ranges. TRD-LL fit} (right) \emph{in the energy range 102.5-109.4 GeV. Signal and background components are represented by the red and blue curves respectively. The magenta line represents the overall fit superimposed on the black data points.}}
   \label{fig:trd}
\end{figure}

The reference spectra for signal and background are evaluated directly from the flight data. For this purpose, pure samples of electrons and protons are selected by means of  tight requirements on the ECAL shower shape, the comparison between reconstructed momentum in the Tracker and measured energy in the ECAL, as well as the reconstructed charge sign. Fig.~\ref{fig:trd} shows the distribution of the TRD-LLe estimator for electrons (left) and protons (center)  in different energy ranges. As expected, the TRD-LLe distribution shows no dependence on the electron energy above $\sim$ 10 GeV. Thus  a unique template function is defined from all electrons selected in the 10-100 GeV energy range and it is used  to represent the signal shape up to the highest energies.  To define the proton template at lower electron energies, the TRD-LLp reference distributions are evaluated separately in each energy interval. The fitting procedure is repeated applying different calorimetric selection of the events. Different cuts on the ECAL BDT estimator are applied to vary the electron purity of the samples and the BDT cut applied in the analysis is chosen such as to minimise the combined systematics and statistical uncertainties. In Fig.~\ref{fig:trd} (right) an example of the fitted signal and background distributions are presented at energies between 102.5 and 109.4 GeV.

In the measurement of the separate electron and positron fluxes, the charge confusion (see Sec.~\ref{sec:posfrac}) has been subtracted after the fitting.

\section{The positron fraction measurement}
\label{sec:posfrac}
The selected sample contains $\sim$6,800,000 primary positrons and electrons and $\sim$700,000 protons. The composition of the sample versus energy is determined by the TRD estimator and $E/p$ matching.

\begin{figure}[h!]
   \centering
   \includegraphics[width=0.5\textwidth]{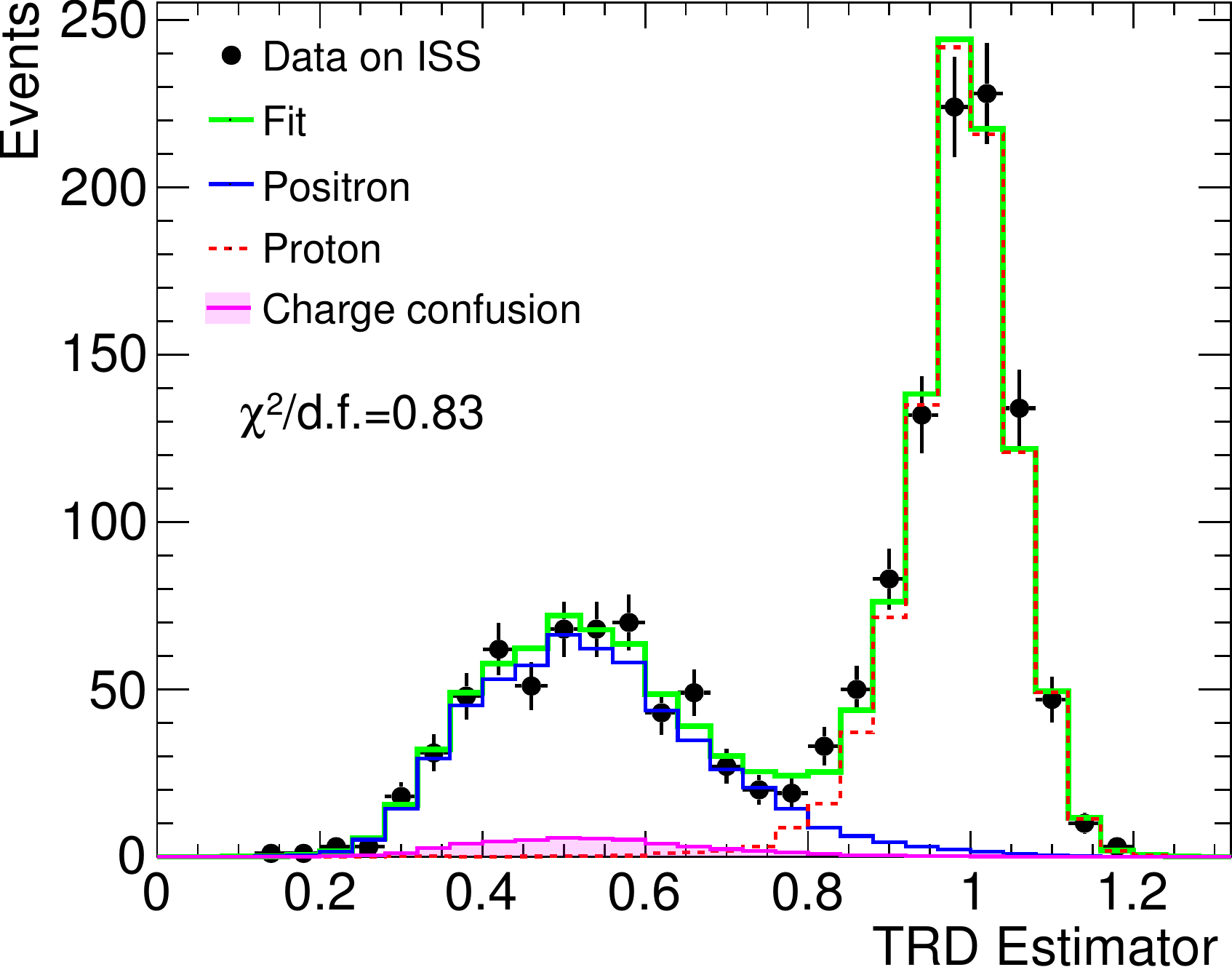}
   \caption{\emph{Separation power of the TRD estimator in the energy range 83.2-100$\,$GeV for the positively charged selected data sample. For each energy bin, the positron and proton reference spectra are fitted to the data to obtain the numbers of positrons and protons.}}
   \label{fig:trdlikelihood}
\end{figure}

In every energy bin, the 2-dimensional reference spectra for e$^\pm$ and the background are fitted to data in the (TRD LL ratio -- $\log(E/p)$) plane by varying the normalizations of the signal and the background. This method provides a data driven control of the dominant systematic uncertainties by combining the redundant TRD, ECAL and Tracker information. The 2D positron reference spectra were verified to be equal to the electron reference spectra using the test beam data. The fit is performed for positive and negative rigidity data samples yielding, respectively, the numbers of positrons and electrons. Results of a fit for the positive sample in the range 83.2--100$\,$GeV are presented in Figure~\ref{fig:trdlikelihood} as a projection onto the TRD estimator axis, where the charge confusion contribution is from electrons misidentified as positrons.

\subsection{Systematic errors}
There are several sources of systematic uncertainty including those associated with the asymmetric acceptance of e$^+$ and e$^-$, the selection of e$^\pm$, bin-to-bin migration, the reference spectra and
charge confusion. The systematic uncertainties were examined in each energy bin over the entire spectrum from 0.5 to 350$\,$GeV.{}

Two sources of charge confusion dominate. The first is related to the finite resolution of the Tracker and multiple scattering. It is mitigated by the $E/p$ matching and the quality cut of the trajectory measurement. The second source is related to the production of secondary tracks along the path of the primary e$^\pm$ in the Tracker. The impact of the second effect was estimated using control data samples of electron events with the ionization in the lower TOF counters corresponding to at least two traversing particles. Both sources of charge confusion are found to be well reproduced by the Monte Carlo simulation. The systematic uncertainties due to these two effects are obtained by varying the background normalizations within the statistical limits. As an example, for the positive sample in the range 83.2--100$\,$GeV the uncertainty on the number of positrons due to the charge confusion is 1.0\,\%.

\section{The fluxes measurement}
\label{sec:measurement}
The electron flux in each energy interval [E, E+$\Delta$ E] is measured as :

$$
\Phi(E,E+ \Delta E)= \frac{N(\!E)}{\Delta E\, A(\!E)\, \Delta T (\!E)\, \epsilon(\!E)}
$$

where $N$ is the number of  electron events; $\Delta$T is the exposure time, 51.6$\times 10^6$ s at energies above 25 GeV; $A$ is the the effective detector acceptance after applying the event selection; $\epsilon$ is the combined efficiency of the trigger and signal selection.

A full MonteCarlo simulation of the response of the AMS-02 detector to an isotropic electron spectrum is used to calculate the detector acceptance. Given the Geometric Factor (GF) of the surface used to generate the events, the acceptance is defined as:

\vspace{-0.1cm}
$$
A(E,E+\Delta E)=GF \times \frac{N_{acc}}{N_{gen}}
$$

where $N_{acc}$ and $N_{gen}$ represent respectively the number of selected and generated events in the energy interval [E, E+$\Delta$E].
The efficiency of each selection cut is evaluated on data and compared to the expectations from simulation.

\subsection{Trigger efficiency}
Different trigger conditions are implemented in the AMS-02 trigger logic to maximize the efficiency for different particle species while keeping a sustainable rate of the recorded events.
Electron events are acquired by either one of the three following conditions:\newline
- {\it single charge trigger} : the coincidence of signals from all the four TOF planes is required in anti coincidence with the ACC ; \newline
- {\it electron trigger}: as for the single charge trigger the four TOF planes coincidence is required, no veto is applied from the ACC signal if energy deposits above threshold are measured in at least two out out of the three ECAL super-layers used in the trigger in both the x and y views; \newline
- {\it photon trigger}: no coincidence of signals from the four TOF planes is found, but there is a shower pointing within the AMS acceptance. A fast reconstruction algorithm is used at the level of the trigger logic to evaluate the shower direction from the signals over threshold registered in the  x and  ECAL super layers views used in the trigger.

In order to measure the trigger efficiency from data, a prescaled sample of events passing looser trigger conditions is also recorded as an {\it unbiased} sample. In particular, 1/100 of the events with a coincidence of signals from at least 3 TOF planes are recorded, irrespectively of any veto from the ACC, and 1/1000 of the events having an energy deposit in the ECAL satisfying the electron trigger condition. \newline The trigger efficiency  is then evaluated from the fraction of electrons selected by the trigger over the total number of electrons in the triggered + {\it unbiased} sample, taking into account the appropriate prescaling factor. Above a few GeVs, no unbiased events were present in the electron sample and the measured efficiency is 100\%. At lower energies, as the energy deposit in the calorimeter decreases, the requirement of the electromagnetic trigger becomes too tight and a reduction of the trigger efficiency is observed due to the effect of the ACC veto.

\subsection{Track reconstruction efficiency}
The efficiency of having a reconstructed track associated to an electron passing through the tracker acceptance has been studied in data  as a function of energy. For this estimate, the same requirements used in the electron analysis flow are applied to the data sample, except the requirement of a track. The efficiency has been defined from the ratio of the number of electrons with an associated track over the  total number of electrons, both quantities are evaluated for particles passing through the geometrical acceptance of the tracker. In Fig.~\ref{fig:stability} (left) the tracker efficiency as a function of the energy is shown. Data and MonteCarlo estimates are in agreement  at the one \% level over a wide energy range.

\subsection{ ECAL BDT selection efficiency}
\begin{figure}[h!]
   \centering
   \includegraphics[width=0.46\textwidth]{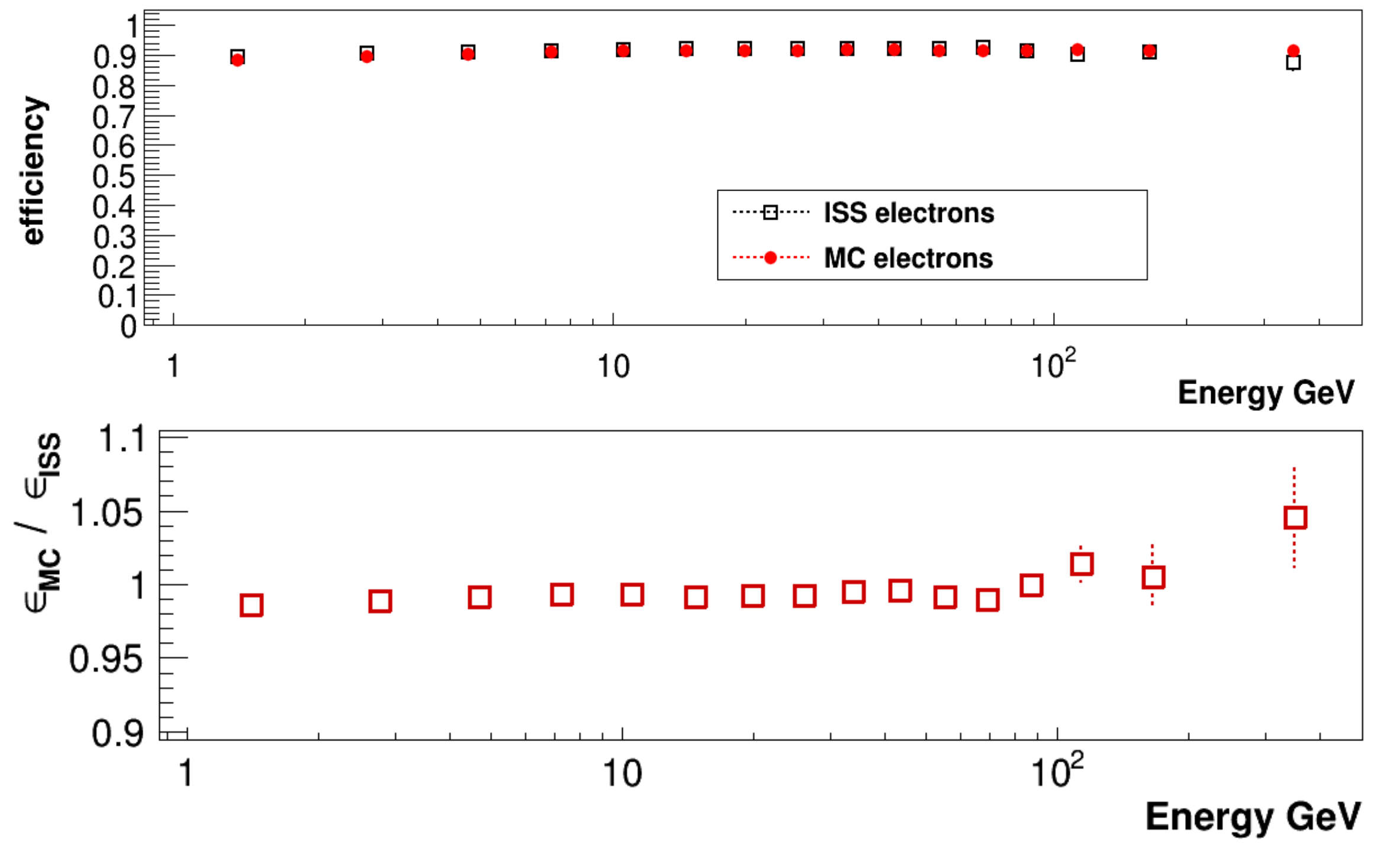}
   \ \
   \includegraphics[width=0.46\textwidth]{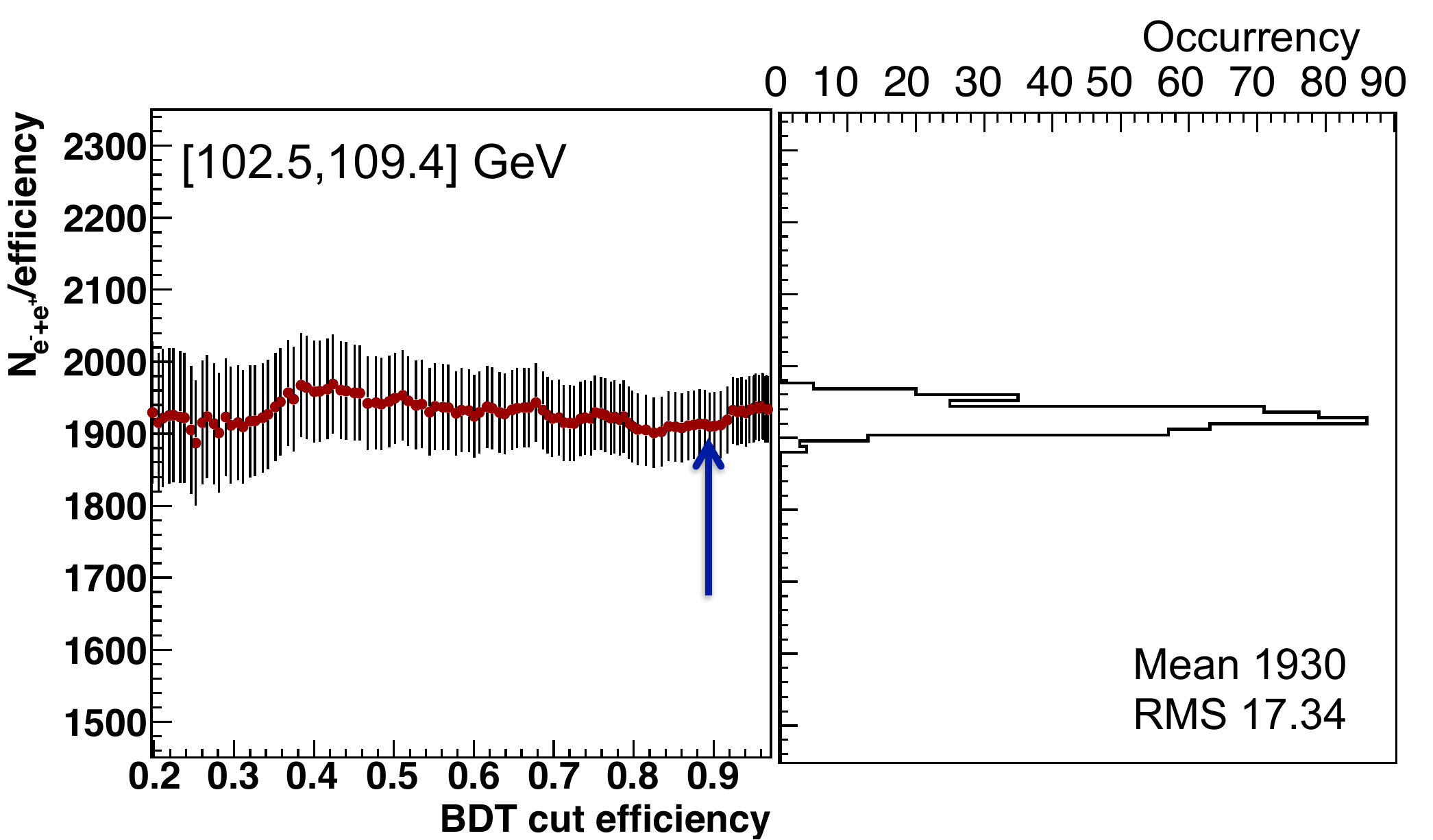}
   \caption{(Left) \emph{Track reconstruction efficiency as a function of energy.} (Right) \emph{Stability of the measured number of electrons as a function of the ECAL BDT cut efficiency. The blue arrow indicates the working point in the presented measurement.}}
   \label{fig:stability}
\end{figure}

In each energy interval the measurement is performed at the ECAL BDT cut that minimizes the overall measurement uncertainties. The ECAL BDT efficiency is evaluated from a probe sample of electrons, chosen with tight requirements on the energy/momentum ratio and a negative charge sign. The template fit analysis is performed on this sample at different BDT cuts and the ratio between fitted electrons at a given BDT cut vs the total number of fitted electrons in absence of calorimetric selection defines the efficiency.
The stability of the measurement against different selection efficiencies  is shown in Fig.~\ref{fig:stability} (right) where the number of electrons corrected by the BDT efficiency is shown as a function of different efficiencies in the BDT cut applied before performing the template fit analysis. The RMS of the distribution is $<$1\% over a wide range leading to a minor contribution to the overall measurement uncertainty.

\section{Conclusions}
\begin{figure}[h!]
   \centering
   \includegraphics[width=0.4\textwidth]{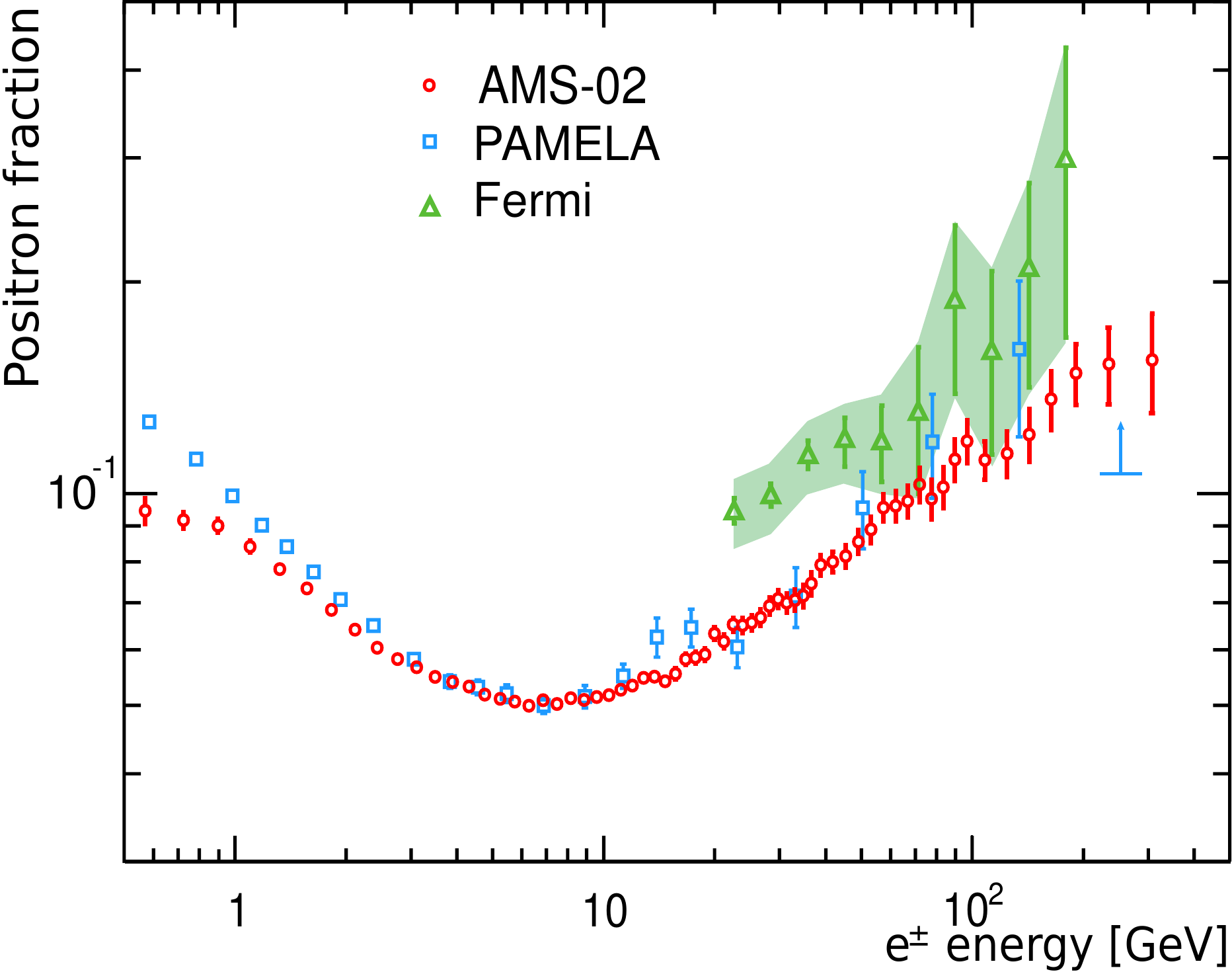}
   \quad
   \includegraphics[width=0.49\textwidth]{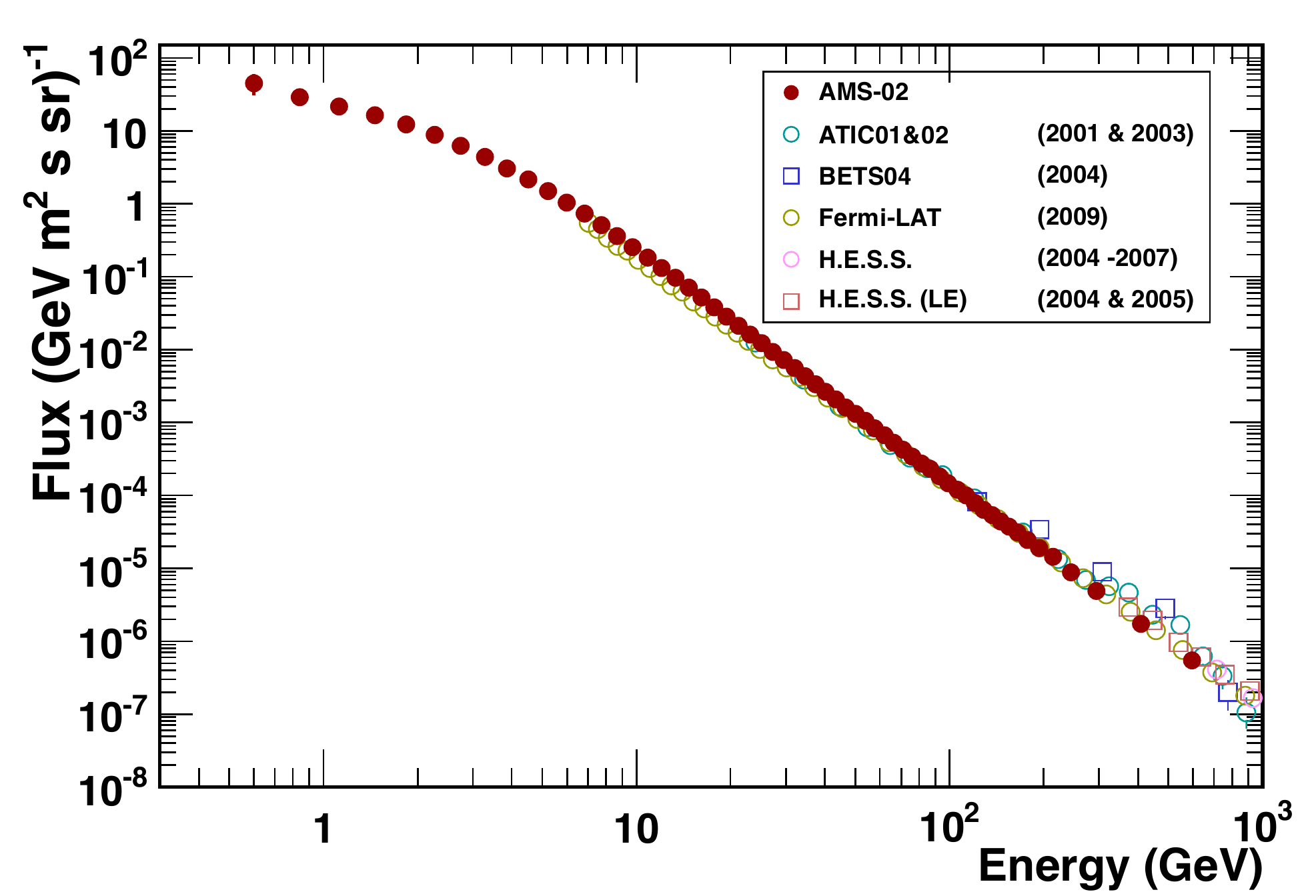} \\
   \includegraphics[width=0.45\linewidth]{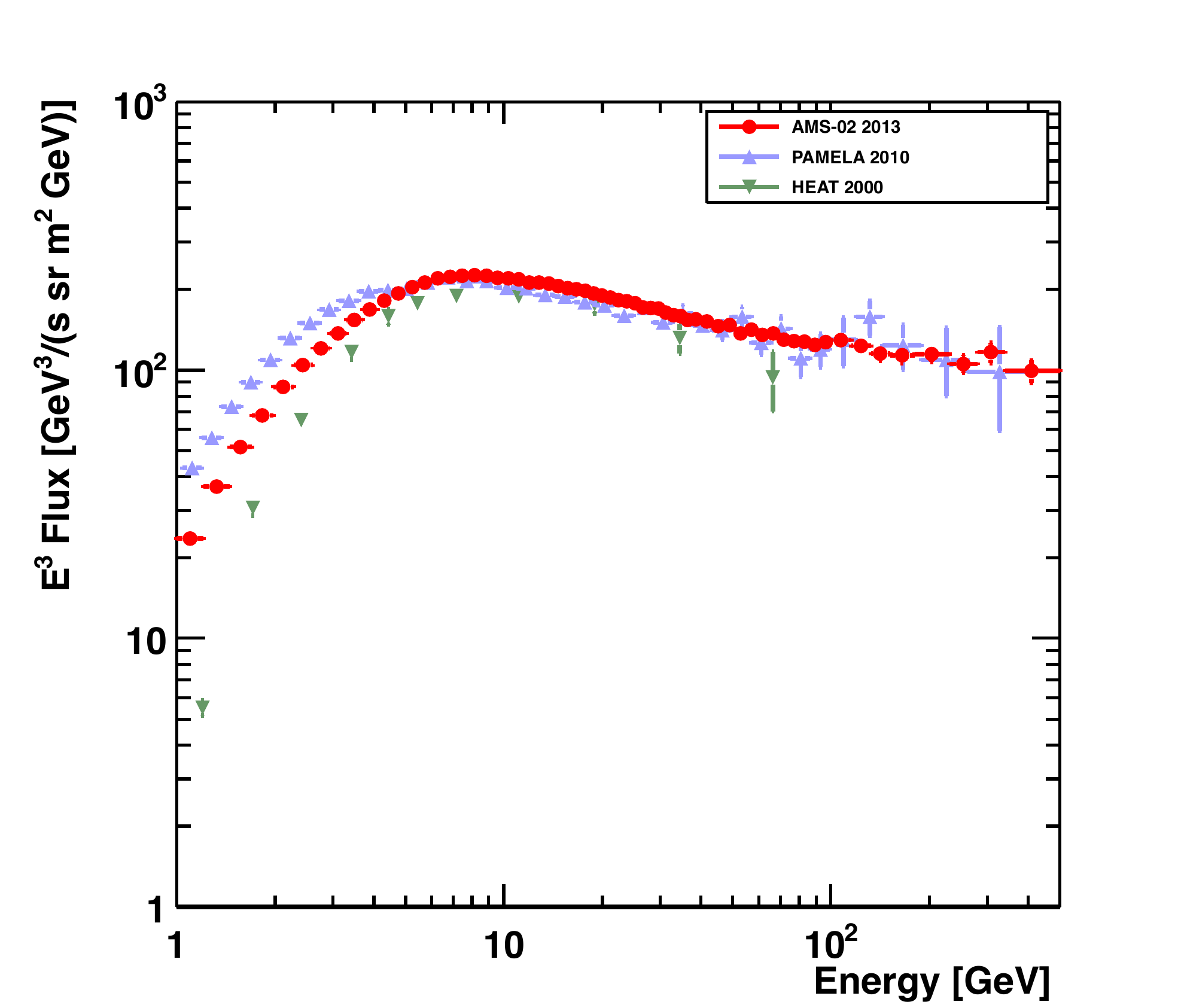}
   \quad
   \includegraphics[width=0.45\linewidth]{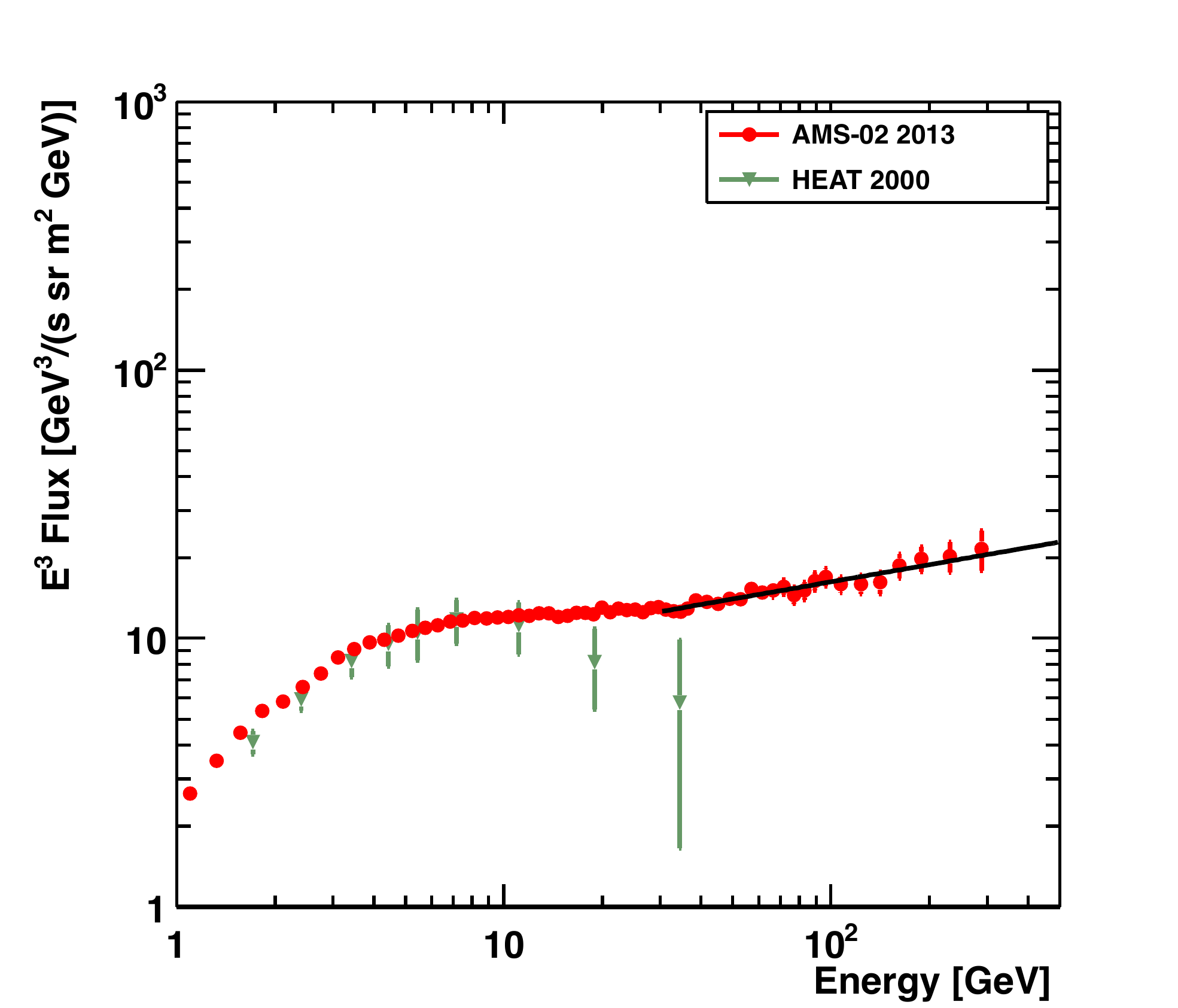}
   \caption{(Left top) \emph{The positron fraction compared with the most recent measurements from PAMELA~\cite{bib15} and Fermi-LAT~\cite{bib16}. The error bars for AMS are the quadratic sum of the statistical and systematic uncertainties and the horizontal positions are the centers of each bin.} (Right top) \emph{AMS combined electrons+positron spectrum (red points) superimposed to recent measurements from references~\cite{bib:ATIC,bib:PPB,bib:FERMI2010,bib:HESS2008,bib:HESS2009}.} (Left and right bottom) \emph{AMS electron and positron spectra compared with the most recent measurements from PAMELA~\cite{pamela} and HEAT~\cite{heat}.}}
   \label{fig:measurement}
\end{figure}

The measurement of the electron spectrum with the AMS-02 detector has been performed at energies between 0.5 and 700 GeV and is reported in Fig.~\ref{fig:measurement} (right top). The assessment of systematic uncertainties is currently being finalised. For this measurement, $\sim$ 9 million electrons have been selected from more than 30 billion trigger collected in two years of operation in space. This represents $\sim$ 10\% of the expected AMS data sample.

The measured positron fraction is presented in Fig.~\ref{fig:measurement} (left top) as a function of the reconstructed energy at the top of the AMS detector. The first 6.8 million primary positron and electron events collected with AMS on the ISS show: at energies $< 10\,$GeV, a decrease in the positron fraction with increasing energy; a steady increase in the positron fraction from 10 to $\sim$250$\,$GeV; the slope of the positron fraction versus energy decreases by an order of magnitude from 20 to 250\,GeV and no fine structure is observed.

Fig.~\ref{fig:measurement} (left and right bottom) show the electron and positron fluxes. The electron flux measurement extends up to 500 GeV. Multiplied by $E^3$ it is rising up to 10 GeV and appears to be on a smooth, slowly falling curve above. The positron flux measurement extends up to 350~GeV. Multiplied by $E^3$  it is rising up to 10 GeV, from 10 to 30 GeV the spectrum is flat and above 30 GeV again rising as indicated by the black line in the figure. The spectral index and its dependence on energy is clearly different from the electron spectrum.

\vspace*{0.3 cm}

\Acknowledgements
This work has been supported by acknowledged person and institutions in~\cite{bib:AMS02-PRL} as well as by the Italian Space Agency  (ASI) under contract ASI-INFN I/002/13/0 and ASDC/011/11/1

\end{document}